\documentclass{eptcs} 
\usepackage{amsmath,amssymb,cite,color,xspace,verbatim}

\newcommand{\tool}{\textsc{InterHorn}\xspace}
\newcommand{\limp}{\rightarrow}
\newcommand{\Land}{\bigwedge}
\newcommand{\ltrue}{\mathit{true}}
\newcommand{\setOf}[1]{\{#1\}}
\newcommand{\vars}{v}
\newcommand{\init}{\mathit{init}}
\newcommand{\next}{\mathit{next}}
\newcommand{\wf}{\mathit{wf}}
\newcommand{\safe}{\mathit{safe}}
\newcommand{\loc}{\ell}

\title{Generalised Interpolation by Solving Recursion-Free 
  Horn Clauses}

\author{Ashutosh Gupta
  \institute{IST, Austria}
  \email{agupta@ist.ac.at}
\and
  Corneliu Popeea
  \institute{Technische Universit\"at M\"unchen}
  \email{popeea@model.in.tum.de}
\and
  Andrey Rybalchenko
  \institute{Technische Universit\"at M\"unchen}
  \institute{Microsoft Research Cambridge}
  \email{rybal@microsoft.com}
}

\def\titlerunning{Generalised Interpolation by Solving Recursion-Free Horn Clauses}
\def\authorrunning{A. Gupta, C. Popeea \& A. Rybalchenko}

\begin{document}
\maketitle

\begin{abstract}

In this paper we present \tool, a solver for recursion-free Horn
clauses. The main application domain of \tool lies in solving
interpolation problems arising in software verification. We show how a
range of interpolation problems, including path, transition, nested,
state/transition and well-founded interpolation can be handled
directly by \tool. By detailing these interpolation problems and their
Horn clause representations, we hope to encourage the emergence of a
common back-end interpolation interface useful for diverse
verification tools.
\end{abstract}

\section{Introduction}

Interpolation is a key ingredient of a wide range of software
verification tools that is used to compute approximations of sets and
relations over program states, see
e.g.~\cite{McMillanCAV06,AlbarghouthiVMCAI12,AlbarghouthiCAV12,WeissenbacherCAV11,SharyginaATVA12,HeizmannPOPL10,POPL11,PLDI12,TACAS12,JaffarCP09}. 
These approximations come in different forms, e.g., as path
interpolation~\cite{JhalaPOPL04}, transition
interpolation~\cite{JhalaCAV05}, nested
interpolation~\cite{HeizmannPOPL10}, state/transition
interpolation~\cite{AlbarghouthiVMCAI12}, or well-founded
interpolation~\cite{SAS05}.
As a result algorithms and tools for solving interpolation problems
have become an important area of research contributing to the advances
in state-of-the-art of software verification.

In this paper we present \tool, a solver for constraints in form of
recursion-free Horn clauses that can be applied on various
interpolation problems occurring in software verification.
\tool takes as input clauses whose literals are either assertions in
the theory of linear arithmetic or unknown relations.
In addition, \tool also accepts well-foundedness conditions on the
unknown relations.
The set of input clauses can represent either a DAG or a tree of
dependencies between interpolants to be discovered.
The output of \tool is either an interpretation of unknown relations
in terms of linear arithmetic assertions that turns the input clauses
into valid implications over rationals/reals and satisfies
well-foundedness conditions, or the statement that no such
interpretation exists.
\tool is sound and complete for clauses without well-foundedness conditions.
(\tool is incomplete when well-foundedness conditions are present,
since it relies on synthesis of linear ranking functions.)
\tool is a part of a general solver for recursive Horn
clauses~\cite{PLDI12} and has already demonstrated its practicability
in a software verification competition~\cite{SVCOMP12}.
The main novelty offered by \tool wrt.\ existing interpolating
procedures~\cite{GriggioTACAS11,PrincessIJCAR10,OpenSmtTACAS10,HoenickeSPIN12}
lies in the ability to declaratively specify the interpolation problem
as a set of recursion-free Horn clauses and the support for
well-foundedness conditions.

% We proceed by illustrating how interpolation problems can be
% represented as recursion-free Horn clauses -- the input to \tool.
% Then we briefly describe the algorithm implemented by \tool as well as
% give some implementation details.

%%% Local Variables: 
%%% mode: latex
%%% TeX-master: "main"
%%% End: 

\section{Interpolation by solving recursion-free Horn clauses}
\label{sec-illustration}

In this section we provide examples of how interpolation related
problems arising in software verification can be formulated as solving
of recursion-free Horn clauses.
This collection of examples is not exhaustive and serves as an illustration of the approach. 
We omit any description of how interpolation is used by verification
methods, since it is out of scope of this paper, and rather focus on
the form of interpolation problems and their representation as
recursion-free Horn clauses.
Further examples can be found in the literature, e.g.,~\cite{PLDI12},
as well as are likely to emerge in the future.

\paragraph{\bf Path interpolation}

Interpolation can be used for the approximation of sets of states
reachable by a program along a given path, see e.g.~\cite{JhalaPOPL04}.
A flat program (transition system) consists of program variables $v$,
an initiation condition $\init(v)$, a set of program
transitions $\setOf{\next_1(v, v'), \dots, \next_N(v,
v')}$, and a description of safe states~$\safe(v)$.
A path is a sequence of program transitions.

Given a path $\next_1(v, v'), \dots, \next_n(v, v')$,
the path interpolation problem is to find assertions $I_0(v), I_1(v),
\dots, I_n(v)$ such that:
\begin{equation*}
  \begin{array}[t]{@{}l@{\qquad}l@{}}
    \init(v) \limp I_0(v), & \\[\jot]
    I_{k-1}(v) \land \next_k(v, v') \limp I_k(v'), 
    & \text{for each $k \in 1..n$}  \\[\jot]
    I_n(v) \limp \safe(v). &
  \end{array}
\end{equation*}
We observe that there are no recursive dependencies induced by the above
implications between the interpolants to be discovered, i.e., $I_0(v)$
does not depend on any other interpolant, while $I_1(v)$ depends on
$I_0(v)$, and $I_n(v)$ depends on $I_0(v), \dots, I_{n-1}(v)$.
\tool leverages such absence of dependency cycles in our solving
algorithm, see Section~\ref{sec-algo}.

\paragraph{\bf Transition interpolation}

Interpolation can be applied to compute over-approximation of program
transitions, see e.g.~\cite{JhalaCAV05}.
Given a path $\next_1(v, v'), \dots, \next_n(v, v')$,
a transition interpolation problem is to find $T_1(v, v'), \dots,
T_n(v, v')$ such that:
\begin{equation*}
  \begin{array}[t]{@{}l@{\qquad}l@{}}
    \next_k(v, v') \limp T_k(v, v'), 
    & \text{for each $k \in 1..n$} \\[\jot]
    \init(v_0) \land T_1(v_0, v_1) \land \dots \land
    T_n(v_{n-1}, v_n) \limp \safe(v_n). & 
  \end{array}
\end{equation*}
Again, we note there are no recursive dependencies between the
assertions to be computed.

\paragraph{\bf Well-founded interpolation}

We can also use interpolation in combination with additional
well-foundedness constraints when proving program termination, see
e.g.~\cite{SAS05}.
We assume a path $\mathit{stem}_1(v, v'), \dots, \mathit{stem}_m(v,
v')$ that contains transitions leading to a loop entry point, and a
path $\mathit{loop}_1(v, v'), \dots, \mathit{loop}_n(v, v')$ 
around the loop. 
A well-founded interpolation problem amounts to finding $I_0(v),
I_1(v), \dots, I_m(v)$, and $T_1(v,v'), \dots, T_n(v,v')$ such that:
\begin{equation*}
  \begin{array}[t]{@{}l@{\qquad}l@{}}
    \init(v) \limp I_0(v), & \\[\jot]
    I_{k-1}(v) \land \mathit{stem}_k(v, v') \limp I_k(v'), 
    & \text{for each $k \in 1..m$}\\[\jot]
    I_m(v) \land \mathit{loop}_1(v, v') \limp T_1(v, v'), & \\[\jot]
    T_{k-1}(v, v') \land \mathit{loop}_k(v', v'') \limp T_k(v, v''),
    & \text{for each $k\in 2..n$} \\[\jot]
    \wf(T_n(v, v')). &
  \end{array}
\end{equation*}
Note that the last clause, which is a unit clause, requires that the
relation $T_n(v, v')$ is well-founded, i.e., does not admit any
infinite chains.

\paragraph{\bf Search tree interpolation}

Interpolation has been used for optimizing the search for solutions
for a constraint programming goal \cite{JaffarCP09}.
In that work, it is considered the case when the search tree
corresponds to the state space exploration of an imperative program
in order to prove some safety property.
A node from the tree is labeled with a formula $s(\vars)$ that is a
symbolic representation for reachable states at a program point.
The tree structure corresponds to program transitions, a node $n$ has
as many children as the transitions starting at the program point
corresponding to $n$~, i.e., $\next_1(\vars,\vars'), \dots,
\next_m(\vars,\vars')$~.
To optimize the search, symbolic states are generalized by computing
interpolants in post-order tree traversal.
During the tree traversal, for a node $n$~, initially labeled $s_0$~,
and having children with labels $s_1$ to $s_m$~, a generalized label
of the node $n$ is computed as $I_1(\vars) \land \dots \land
I_m(\vars)$ and is subject to the following implications:

\begin{equation*}
  \begin{array}[t]{@{}l@{\qquad}l@{}}
    s_0(\vars) \limp I_1(\vars) \land \dots \land I_m(\vars) & \\[\jot]
    I_k(\vars) \limp (\next_k(\vars,\vars') \limp s_k(\vars')) 
    & \text{for each $k \in 1..m$}
  \end{array}
\end{equation*}

\noindent
These implications correspond to the following recursion-free Horn
clauses,

\begin{equation*}
  \begin{array}[t]{@{}l@{\qquad}l@{}}
    s_0(\vars) \limp I_k(\vars), & \text{for each $k \in 1..m$} \\[\jot]
    I_k(\vars) \limp (\exists \vars': \next_k(\vars,\vars') \limp s_k(v')),
    & \text{for each $k \in 1..m$}
  \end{array}
\end{equation*}

\noindent
where the quantifier elimination in $\exists \vars':
\next_k(\vars,\vars') \limp s_k(v')$ can be automated for $\next_k$
and $s_k$ background constraints in the theory of linear arithmetic.

\iffalse
\paragraph{\bf Example for search tree interpolation}

We use as example the program from Figure~2 of \cite{JaffarCP09}.
The program has one variable $x$.
One interpolation problem obtained from the search tree traversal
algorithm \cite{JaffarCP09} corresponds to the following set of Horn
clauses:
%
\begin{equation*}
  \begin{array}[t]{ll}
    x=0 \limp I_{D1}(\vars) \land I_{D2}(\vars), \\[\jot]
    I_{D1}(\vars) \limp (\exists \vars': x'=x \limp x'\leq 7), \\[\jot]
    I_{D2}(\vars) \limp (\exists \vars': x'=x+4 \limp x'\leq 7),
  \end{array}
\end{equation*}

\noindent
where the formula $\mathit{node}(D) = (x=0)$ appears in the body of
the first clause, while the formulas $\mathit{node}(F) = (x \leq 7)$
and $\mathit{node}(G) = (x \leq 7)$ appear in the head of the second
and respectively third clause.
Our implementation produces different solutions than those optimal for
optimizing the search tree traversal, however we note that the
interpolation problems obtained from the algorithm \cite{JaffarCP09}
are expressible as recursion-free Horn clauses and solvable by \tool.
\fi

\paragraph{\bf Nested interpolation}

For programs with procedures, interpolation can compute
over-approximations of sets of program states that are expressed over
variables that are in scope at respective program locations, see
e.g.~\cite{JhalaPOPL04,HeizmannPOPL10}.
A procedural program consists of a set of procedures $P$ including the
main procedure $\mathit{main}$, global program variables $g$ that
include a dedicated variable for return value passing, as well as
procedure descriptions.
For each procedure $p\in P$ we provide its local variables $l_p$, a
finite set of intra-procedural program transitions of the form
$\mathit{inst}^p(g, l_p, g', l'_p)$, a finite set of call transitions
of the form $\mathit{call}^{p, q}(g, l_p, l_q)$ where $q\in P$ is the
name of the callee, a finite set of return transitions of the
form~$\mathit{ret}^p(g, l_p, g')$, as well as a description of safe
states~$\safe^p(g, l_p)$.

A path in a procedural program is a sequence of program transitions
(including intra-procedural, call and return transitions) that
respects the calling discipline, which we do not formalize here.

Given a path $\next_1(v, v'), \dots, \next_n(v, v')$. 
Find $I_0(v_0), I_1(v_1), \dots, I_n(v_n)$, where $v_0, \dots, v_n$
are determined through the following implications, such that: 
\begin{equation*}
  \begin{array}[t]{@{}l@{}}
    \init(g, l_\mathit{main}) \limp I_0(g, l_\mathit{main}),
    \\[\jot]
    \begin{array}[t]{@{}l@{\;}l@{}}
    I_{k-1}(g, l_p) \mathrel{\land}
%    \\[\jot]   \quad
    \begin{cases}
      \begin{array}[t]{@{}l@{\;}l@{}}
        \mathit{inst}^p(g, l_p, g', l'_p) \limp I_k(g', l'_p),
        &
        \text{if $\next_k(v, v') = \mathit{inst}^p(g, l_p, g',
          l'_p)$}
        \\[\jot]
        \mathit{call}^{p,q}(g, l_p, l_q) \limp I_k(g, l_q),
        &
        \text{if $\next_k(v, v') = \mathit{call}^{p,q}(g, l_p,
          l_q)$}
        \\[\jot]
        \mathit{ret}^p(g, l_p, g') \limp I_k(g', l_q),
        &
        \begin{array}[t]{@{}l@{}}
          \text{if $\next_k(v, v') = \mathit{ret}^p(g, l_p,
            g')$ returns to $q$} \\[\jot]
        \end{array}
      \end{array}
    \end{cases}\\
  \end{array}\\
  \text{for each $k\in 1..n$}\\[\jot]
    I_n(g, l_p) \limp \safe^p(g, l_p), 
    \quad
    \text{when $\next_n(v, v')$ occurs in procedure $p$. }
  \end{array}
\end{equation*}
Similarly to the previously described interpolation problems, there
are no recursive dependencies in the above clauses. 

\paragraph{\bf State/transition interpolation}

As illustrated by the example of well-founded interpolation,
interpolants can represent over-approximations of sets of states as
well as binary relations.
The Whale algorithm provides a further example of such
usage~\cite{AlbarghouthiVMCAI12}.
Given a sequence of assertions $\next_1(v, v'), \dots,
\next_n(v, v')$ that represent an under-approximation of a
path through a procedure with a guard $\mathit{g}(v)$ and a summary
$\mathit{s}(v, v')$.
Find guards $G_1(v), \dots, G_n(v)$ and summaries $S_1(v, v'), \dots,
S_n(v, v')$ such that:
\begin{equation*}
  \begin{array}[t]{@{}l@{\qquad}l@{}}
    \next_k(v, v') \limp S_k(v, v'), 
    & \text{for each $k \in 1..n$} \\[\jot]
    g(v) \limp G_1(v), & \\[\jot]
    G_k(v) \land S_k(v, v') \limp G_{k+1}(v'), 
    & \text{for each $k\in 1..n-1$}
    \\[\jot]
    G_n(v) \land S_n(v, v') \limp s(v, v'). & 
  \end{array}
\end{equation*}
There are no recursive dependencies among the unknown guards and
summaries.

\paragraph{\bf Solving unfoldings of recursive Horn clauses}

A variety of reachability and termination verification
problems for programs with procedures, multi-threaded programs, and
functional programs can be formulated as the satisfiability of a set
of recursive Horn clauses, e.g.,~\cite{CAV11,POPL11,PLDI12}.
These clauses are obtained from the program during a so-called
constraint generation step.
The satisfiability checking performed during the constraint solving
step amounts to the inference of inductive invariants, procedure
summaries, function types and other required auxiliary assertions.
Existing solvers, e.g., HSF~\cite{PLDI12} and
$\mu$\textit{Z}~\cite{GenPDRSAT12}, rely on solving recursion-free
unfoldings when iteratively constructing a solution for recursive Horn
clauses.

We illustrate the generation of recursion-free unfolding using an
invariance proof rule for flat programs.
This rule can be formalised by as follows.
For a given program find an invariant $\mathit{Inv}(v)$ such that
\begin{equation*}
  \begin{array}[t]{@{}l@{}}
    \init(v) \limp \mathit{Inv}(v), \\[\jot]
    \mathit{Inv}(v) \land \next(v, v')\limp \mathit{Inv}(v'), 
    \quad \text{for each program transition $\mathit{next(v, v')}$}\\[\jot]
    \mathit{Inv}(v) \limp \safe(v).
  \end{array}
\end{equation*}
An unfolding of these recursive clauses introduces auxiliary relations
that refer to $\mathit{Inv}(v)$ at each intermediate step. 
For example we consider an unfolding that starts with the first clause
above and then applies a clause from the second line for a transition
$\next_1(v, v')$ and then for a transition $\next_2(v,
v')$ before traversing the last clause.
This unfolding is represented by the following recursion-free clauses:
\begin{equation*}
  \begin{array}[t]{@{}l@{}}
    \init(v) \limp \mathit{Inv}_0(v),\;
    \mathit{Inv}_0(v) \land \next_1(v, v')\limp\mathit{Inv}_1(v'),\\[\jot]
    \mathit{Inv}_1(v) \land \next_2(v, v')\limp \mathit{Inv}_2(v'),\;
    \mathit{Inv}_2(v) \limp \safe(v). 
  \end{array}
\end{equation*}
A solution for these clauses contributes to solving the recursive clauses.

%%% Local Variables: 
%%% mode: latex
%%% TeX-master: "main"
%%% End: 

\section{Algorithm overview}
\label{sec-algo}

In this section we briefly describe how \tool solves recursion-free
Horn clauses. 
We refer to \cite[Section~7]{POPL11} for a solving algorithm for
clauses over linear rational arithmetic, to \cite{APLAS11} for a treatment
of a combined theory of linear rational arithmetic and uninterpreted
functions, and to \cite{TACAS12} for a support of well-foundedness
conditions.

\tool critically relies on the following two observations. 
First, applying resolution on clauses that describe the interpolation problem
terminates and yields an assertion that does not contain any unknown
relations.
For example, resolution of clauses in Section~\ref{sec-illustration}
that describe path, transition, nested and state/transition
interpolation results in the implication of the form
$\mathit{init}(v_0) \land (\Land_{k=1}^n \mathit{next}_k(v_{k-1},
v_k)) \limp \mathit{safe}(v_n)$.
Second, the obtained assertion is valid if and only if the set of
clauses is satisfiable.
From the proof of validity (or alternatively, from the proof of
unsatisfiability of the negated assertion) we construct the solutions.

\paragraph{\bf Clauses without well-foundedness conditions}

\tool goes through three main steps when given a set of recursion-free
clauses that does not contain any well-foundedness condition.
For example, we consider the following recursion-free clauses as input:
\begin{equation*}
x \geq 10 \limp p(x),\;\;
  p(u) \land w = u+v \limp q(v, w),\;\;
  q(y, z) \land y \leq 0 \limp z \geq y.
\end{equation*}

During the first step we apply resolution on the set of clauses.
Since the clauses are recursion-free, the resolution application
terminates. 
The result is an assertion that only contains constraints from the
background theory. 
After applying resolution we obtain for our example (note that we use
fresh variables here to stress the fact that clauses are implicitly
universally quantified):
$a \geq 10 \land c = a+b \land b \leq 0 \limp c \geq b$.

The second step amounts to checking the validity of the obtained
assertion.\footnote{Instead of validity checking we can check
satisfiability of the negated assertion.}
If the assertion is not valid then we report that the original set of
clauses imposes constraints that cannot be satisfied.
Otherwise we produce a proof of validity.
In our example the proof of validity can be represented as a weighted
sum of the inequalities in the antecedent of the implication, with the
weights 1, $-1$, and 0, respectively. 

The third step traverses the input clauses and computes the solution
assignment by taking the proof into account.
For the clause $x\geq 10 \limp p(x)$ we determine that $x \geq 10$
contributes to $p(x)$ with a weight 1, since during the resolution $x
\geq 10$ gave rise to $a \geq 10$ whose weight is~1.
Thus we obtain $p(x) = (x \geq 10)$.
For the clause $p(u) \land w = u +v \limp q(v, w)$ we combine $p(u)$
and $w = u +v$ with the weight of the latter set to $-1$, since $w = u
+v$ yielded a contribution to the proof with weight~$-1$.
This leads to $q(v, w) = (u\geq10)+(-1)*(w=u+v)=(w\geq 10+v)$.

Finally, \tool outputs the solution:

\begin{equation*}
p(x) = (x \geq 10),\;\; q(v,w) = (w \geq 10+v).
\end{equation*}
We observe that the substitution of the solutions into the input
clauses produces valid implications: $x \geq 10 \limp x \geq 10,$ $u
\geq 10 \land w = u+v \limp w \geq 10+v,$ and $z \geq 10+y \land y \leq 0
\limp z \geq y$.

\paragraph{\bf Clauses with a well-foundedness condition}
In case of a well-foundedness condition occurring in the input, \tool
introduces additional steps to take this condition into account.
For example, we consider the following recursion-free clauses with a
well-foundedness condition as input:
\begin{equation*}
  \begin{array}[t]{@{}c@{}}
    x \geq 10 \limp p(x),\quad
    p(u) \land w = u+v \limp q(v, w),\quad
    q(y, z) \land y \leq 0 \limp r(y, z), \\[\jot]
    \mathit{wf}(r(s, t)).
  \end{array}
\end{equation*}

The first step is again the resolution of the given clauses that
produces a clause providing an under-approximation for the relation
that is subject to the well-foundedness condition.
For our example, we obtain: 
$a \geq 10 \land c = a+b \land b \leq 0 \limp r(b, c)$.

The second step attempts to find a well-founded relation that
over-approximates the projection of the antecedent of the clause
obtained by resolution on the variables in its head.
For our example this projection amounts to performing an
existential quantifier elimination on $\exists a: a \geq 10 \land c
= a+b \land b \leq 0$, which gives~$c \geq 10+b \land b \leq 0$.
This relation is well-founded, which is witnessed by a ranking
relation over $b$ and $c$ with a bound component $b \leq 0$ and the
decrease component $c \geq b+1$.

The third step uses the well-founded over-approximation to construct a
clause that introduces an upper bound on the relation under
well-foundedness condition.
This clause replaces the well-formedness condition by an approximation
condition wrt.\ an assertion.
For our example, the clause $\mathit{wf}(r(s,t))$ is replaced by the
clause $r(s, t) \limp (s \leq 0 \land t \geq s+1)$.

Lastly, we apply the solving method for clauses without
well-foundedness conditions described previously. In our example, the
set of clauses to be solved becomes:
\begin{equation*}
  \begin{array}[t]{@{}c@{}}
    x \geq 10 \limp p(x),\quad
    p(u) \land w = u+v \limp q(v, w),\quad
    q(y, z) \land y \leq 0 \limp r(y, z), \\[\jot]
    r(s, t) \limp (s \leq 0 \land t \geq s+1).
  \end{array}
\end{equation*}

Finally, \tool outputs the solution:

\begin{equation*}
p(x) = (x\geq10),\;\; q(v,w) = (w\geq10+v),\;\; r(s,t) = (s\leq0 \land t\geq s+10).
\end{equation*}

\section{Implementation}

\tool is implemented in SICStus Prolog~\cite{sicstus}. 
For computing proofs of validity (resp.\ unsatisfiability) over linear
rational arithmetic theory, \tool relies on a proof producing version
of a simplex algorithm~\cite{GuptaPhD}.
For computing well-founded approximations (also over linear rational
arithmetic theory), \tool uses a linear ranking functions synthesis
algorithm~\cite{VMCAI04}.
\tool can be downloaded from 
\url{http://www7.in.tum.de/tools/interhorn/},
accepts input in form of Prolog terms 
and outputs an appropriately formatted result.

%%% Local Variables: 
%%% mode: latex
%%% TeX-master: "main"
%%% End: 

%\input{examples}

\section{Conclusion}

We presented \tool, a solver for recursion-free Horn clauses that can
be used to deal with various interpolation problems. 
The main directions for the future development include adding support
for uninterpreted functions, along the lines of~\cite{APLAS11}, and
integer arithmetic.
After developing our work, we became aware of a related work
highlighting the relation between interpolation and recursion-free
Horn clauses \cite{RummerVSTTE13}.
The authors of \cite{RummerVSTTE13} show that some interpolation
problems correspond to various fragments of recursion-free Horn
clauses and establish complexity results for these fragments assuming
the background theory of linear integer arithmetic.
Our work is less concerned with the different fragments of
recursion-free Horn clauses and more with how interpolation problems
arise in software verification.
The well-founded interpolation problem is beyond the scope of
\cite{RummerVSTTE13}.

\subsubsection*{Acknowledgements}

This research was supported in part by ERC project 308125 VeriSynth,
by Austrian Science Fund NFN RiSE (Rigorous Systems Engineering),
and by the ERC Advanced Grant QUAREM (Quantitative Reactive Modeling).

\bibliographystyle{eptcs}
\bibliography{biblio}

\end{document}